\def\year{2019}\relax
\documentclass[letterpaper]{article} 
\usepackage{aaai19}  
\usepackage{times}  
\usepackage{helvet} 
\usepackage{courier}  
\usepackage{graphicx} 
\usepackage[hyphens]{url}  
\usepackage{graphicx}  
\usepackage{balance}
\usepackage{booktabs}

\usepackage{xspace}
\newcommand{\BfPara}[1]{{\noindent\bf#1.}\xspace}
\usepackage{tikz,pgf}

\newcommand*\cib[1]{\tikz[baseline=(char.base)]{
                            \node[shape=circle,fill=blue!50!black,text=white,draw,inner sep=0.3pt] (char) {#1};}}
                            
\title{W-Net: A CNN-based Architecture for White Blood Cells Image Classification}
\author{
    \large{\textbf{Changhun Jung\textsuperscript{\rm 1} Mohammed Abuhamad\textsuperscript{\rm 1,2} Jumabek Alikhanov\textsuperscript{\rm 1}}}\\
    \large{\textbf{Aziz Mohaisen\textsuperscript{\rm 2} Kyungja Han\textsuperscript{\rm 3} DaeHun Nyang\textsuperscript{\rm 1}}}
    \\
    \textsuperscript{\rm 1}Inha University, \textsuperscript{\rm 2}University of Central Florida, \textsuperscript{\rm 3}The Catholic University of Korea\\ 
    jcptk677@gmail.com, abuhamad@knights.ucf.edu, Jumabek4044@gmail.com,
    \\mohaisen@cs.ucf.edu, hankja@catholic.ac.kr, nyang@inha.ac.kr 
}

\begin{document}

\maketitle

\begin{abstract}
Computer-aided methods for analyzing white blood cells (WBC) have become widely popular due to the complexity of the manual process. 
Recent works have shown highly accurate segmentation and detection of white blood cells from microscopic blood images. However, the classification of the observed cells is still a challenge and highly demanded as the distribution of the five types reflects on the condition of the immune system. 
This work proposes W-Net, a CNN-based method for WBC classification. 
We evaluate W-Net on a real-world large-scale dataset, obtained from The Catholic University of Korea, that includes 6,562 real images of the five WBC types.
W-Net achieves an average accuracy of 97\%.
\end{abstract}\vspace{-3mm}

\section{Introduction}
White blood cells (WBCs) are one type of blood cells, besides red blood cell and platelet, and are responsible for the immune system, protecting against foreign substances and bacteria. WBCs are categorized into five major subtypes: neutrophils, eosinophils, basophils, lymphocytes and monocytes. Neutrophils include two functionally unequal subpopulations: neutrophil-killers and neutrophil-cagers, and they defend against bacterial or fungal infections. Eosinophils rise in response to allergies, parasitic infections, collagen diseases, and disease of the spleen and central nervous system. Basophils are chiefly responsible for allergic and antigen response by releasing chemical histamine causing the dilation of blood vessels. Lymphocytes help immune cells to combine with other foreign invasive organisms such as microorganisms and antigens, in order to remove them out of the body. Monocytes phagocytose foreign substances in the tissues. The distribution of these five classes is 62\%, 2.3\%, 0.4\%, 30\% and 5.3\% among WBCs in the body.

Leukemia is a disease in which immature WBCs in the blood abnormally proliferate, rapidly decreasing the number of normal blood cells and making the immune system vulnerable to infection and even leading to death. In the US, about 60,000 people are diagnosed with leukemia every year, and about 20,000 people die of leukemia annually. From 2011 to 2015, leukemia was the sixth most common cause of cancer-caused death in the US~\cite{LeukemiaStatisticsA}. There are various types of leukemia, including ALL (Acute lymphocytic leukemia), AML (Acute myelogenous leukemia), CLL (Chronic lymphocytic leukemia), CML (Chronic myelogenous leukemia). Chronic leukemia progresses more slowly than acute leukemia which requires immediate medical care. Acute leukemia is characterized by proliferation of blasts, CLL is characterized by increased lymphocytes while CML shows markedly increased neutrophils and some basophils in the blood~\cite{TypesOfLeukemia}. It is therefore important to analyze the count of the five types of WBC, which will help accurately diagnose leukemia.

In this paper, we propose a CNN-based WBC classification model, W-Net, to accurately recognize WBC types. Our model consists of three convolutional layers for extracting features from WBC images, and two fully-connected layers for classifying them into five classes using a softmax classifier. W-Net outperforms state-of-the-art schemes in terms of accuracy. We compare W-Net to ResNet~\cite{he2016deep} to show its effectiveness in WBC image classification. We further show experiment the LISC public data~\cite{rezatofighi2011automatic} to show how other researchers can benefit from our trained W-Net model.

\BfPara{Contribution} The contribution of this work is as follows. 1) We propose a CNN architecture, W-Net, with a small number of layers fostering efficiency for WBC classification. 2) We examine the performance of W-Net using a large-scale dataset consisting of 6,562 real images. 3) We address an unbalanced dataset for five classes with deep learning and obtain a WBC classification accuracy of 97\%.

\BfPara{Organization} The rest of this paper covers a review of the literature, the W-Net model, an evaluation through various experiments on WBC images, and concluding remarks.

\section{Related Work} \label{sec:related_work}

\begin{table*}[t]
\centering
\renewcommand{\arraystretch}{1}
\caption{Related work highlighting the used datasets, their size, number of classes (C), employed methods, and accuracy.}
\vspace{2mm}
\label{table:related_works} 
{\footnotesize
{
\begin{tabular}{p{0.21\linewidth}p{0.19\linewidth}p{0.03\linewidth}p{0.01\linewidth}p{0.28\linewidth}p{0.07\linewidth}}
\toprule

\multicolumn{1}{c}{\textbf{Study}} & \multicolumn{1}{c}{\textbf{Dataset}} & \multicolumn{1}{c}{\textbf{Size}} & \multicolumn{1}{c}{\textbf{C}} & \multicolumn{1}{c}{\textbf{Methods}} & \multicolumn{1}{c}{\textbf{Accuracy}} \\

\midrule

\cite{wang2016spectral} & Private & --- & 5 & Morphology, spectral analysis, SVM & 90.00\% \\

\cite{dorini2012semiautomatic} & CellAtlas & 100 & 5 & Morphological Transform., KNN & 78.51\% \\



\cite{hegde2018development} & Private & 117 & 5 & Arithmetical operations, ANN & 96.50\% \\
\cite{ghosh2016blood} & Private & 150 & 5 & Segmentation, morphology, TFC & --- \\
\cite{rawat2018application} & Private~\cite{WBCnucleidataset} & 160 & 4 & Ensemble ANN & 95.00\% \\
\cite{mathur2013scalable} & Private & 237 & 5 & NB & 92.72\% \\
\cite{nazlibilek2014automatic} & Kanbilim~\cite{Kanbilim} & 240 & 5 & Thresholding, ANN, PCA & 95.00\% \\
\cite{putzu2014leucocyte} & ALL-IDB & 260 & 2 & SVM & 92.00\% \\
\cite{abdeldaim2018computer} & ALL-IDB2 & 260 & 2 & Thresholding, KNN, SVM, NB, DT & 91.67\% \\
\cite{ghosh2017simultaneous} & ALL-IDB & 260 & 2 & CNN & 97.22\% \\
\cite{ramesh2012isolation} & Private & 320 & 5 & LDA & 93.90\% \\
\cite{habibzadeh2018automatic} & Personal~\cite{WBCclassification} & 352 & 4 & CNN & 93.17\% \\
\cite{liang2018combining} & BCCD~\cite{BCCDdataset} & 364 & 4 & RNN (LSTM), CNN & 90.79\% \\
\cite{su2014neural} & \cite{CellaVision} & 450 & 5 & Morphology, NN, SVM, MLP & 95.18\% \\
W-Net (this work) & Private & 6,562 & 5 & CNN & 97.00\%\\
W-Net (this work) & LISC public data & 254 & 5 & CNN, further training & 96.00\%\\
\bottomrule

\end{tabular}}}\vspace{-4mm}
\end{table*}

The analysis of WBCs is important for diagnosing diseases. The distribution of the WBC types reflects the condition of the immune system. Analyzing WBCs's components requires performing a segmentation and classification process.  

The conventional process of analyzing WBCs includes the observation of a blood smear through a microscope and the classification process relies on visible characteristics such as shape and color. However, the accuracy of WBCs analysis depends significantly on the knowledge and experience of the medical operator~\cite{wang2016spectral}, while being  time-consuming and labor-intensive~\cite{wang2016spectral,andrade2019recent}. Thus, computer-aided methods have been introduced to enable accurate analysis for the segmentation and identification of WBCs, and to replace the manual method when it is not needed. 

Shitong and Min~\cite{shitong2006new} proposed an algorithm based on fuzzy cellular neural networks to detect WBCs in microscopic blood images as the first key step for automatic WBC recognition. Using mathematical morphology and fuzzy cellular neural networks, they achieved a detection accuracy of 99\%. The detection of WBCs is followed by  segmenting an image into nucleus and cytoplasm regions. This task has been pursued by several studies providing accurate segmentation using a variety of methods. The most common approach for nuclei segmentation is clustering based on the extracted features from pixel values~\cite{andrade2019recent,jiang2006novel,viswanathan2015fuzzy,alferez2015automatic,moradiamin2016computer}. The literature shows a successful nuclei segmentation using different clustering techniques, such as K-means~\cite{gautam2014white}, fuzzy K-means~\cite{moradiamin2016computer,alferez2015automatic,viswanathan2015fuzzy}, C-means~\cite{viswanathan2015fuzzy}, and GK-means~\cite{mohapatra2011fuzzy}. 

Other studies utilized thresholding~\cite{nazlibilek2014automatic,tosta2015unsupervised,abdeldaim2018computer,cao2018novel,mohammed2013chronic}, arithmetical operations~\cite{hegde2018development}, edge-based detection~\cite{viswanathan2015fuzzy,mohammed2013chronic}, region-based detection~\cite{mohammed2013chronic}, genetic algorithms~\cite{chan2010leukocyte}, watershed algorithms~\cite{jiang2006novel}, and Gram-Schmidt orthogonalization~\cite{rezatofighi2011automatic}. The literature on WBCs segmentation is rich and provides valuable insights for WBCs identification. Andrade et al.~\cite{andrade2019recent} provide a survey and a comparative study of 15 segmentation methods using five public WBC databases. Some of these works are dedicated to  adjacent cells' separation, while others addressed overlapping cells.

After segmentation, the WBC image classification or identification is conducted. The difference between WBCs identification and WBC image classification is that the identification process aims to detect and identify Leucocytes in an image, while the classification process aims to distinguish the different types of WBC. Although many studies are dedicated to segmentation and identification task, fewer addressed the classification of WBCs.The literature shows that classification methods used for this purpose include the K-Nearest Neighbor (KNN) classifier~\cite{dorini2012semiautomatic,abdeldaim2018computer,chatap2014analysis}, Bayesian classifier~\cite{mathur2013scalable,abdeldaim2018computer}, Support Vector Machine (SVM) classifier~\cite{wang2016spectral,abdeldaim2018computer,mohapatra2011fuzzy,jagadev2017detection,su2014neural,rezatofighi2011automatic}, Linear Discriminant Analysis (LDA)~\cite{ramesh2012isolation}, decision trees and random forest classifier~\cite{ghosh2016blood,abdeldaim2018computer}, and deep learning~\cite{huang2018convolutional,nazlibilek2014automatic,hegde2018development,rezatofighi2011automatic,habibzadeh2018automatic,rawat2018application,su2014neural}. Table~\ref{table:related_works} summarizes the related work.

\section{W-Net}

\begin{figure*}[!t]
	\centering
	\includegraphics[scale=0.56]{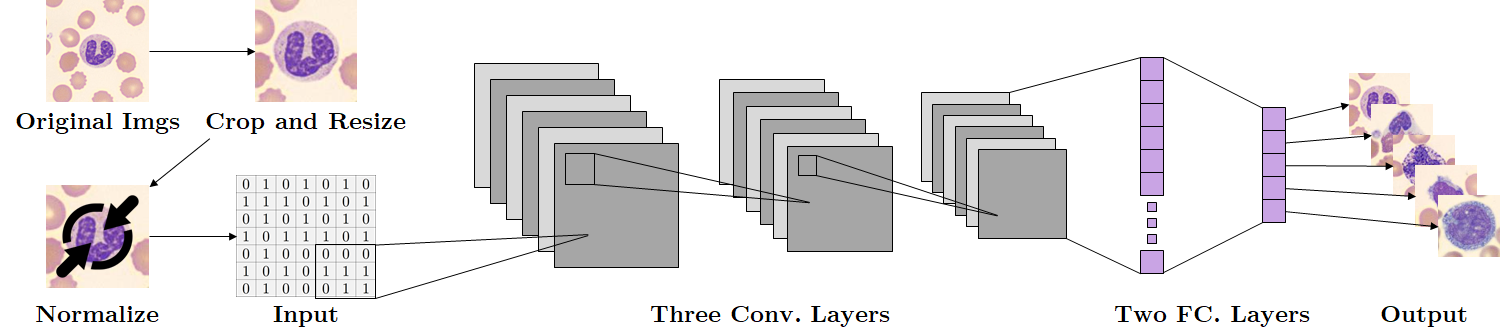}
	\caption{
		An overview of the pre-processing and the proposed CNN-based architecture for WBC image classification. The pre-processing consists of cropping, re-sizing and normalizing. Three convolutional layers (including three pooling layers) are in charge of extracting and learning features, and two fully connected layers are in charge of classification.
	}~\label{fig:w-net_network_img}\vspace{-5mm}
\end{figure*}

In this section, we introduce our CNN-based architecture, W-Net, for WBC image classification. As illustrated in Figure~\ref{fig:w-net_network_img}, the proposed network consists of three convolutional layers for extracting and learning features, and two fully connected layers for classification. Each convolutional layer has a kernel size of 3$\times$3 with stride of size 1 and uses ReLU activation function and Xavier initializer. The first convolutional layer has 16 filters, the second has 32 filters and third has 64 filters. After each convolutional layer, there is a max-pooling layer of size 2$\times$2 with stride of size 2 and zero padding. We also use dropout with $p=0.6$ to prevent overfitting in each convolutional layer. The output of the third convolutional layer is flattened and fed into the first fully connected layer which has 1024 units. ReLU activation, and dropout with $p=0.6$ are followed. The second fully connected layer has five units (five classes of WBC) and is followed by softmax classifier to map the output (features) to one of the five classes. The network has a total size of 16,806,949 trainable parameters. For the training, we use the softmax loss function, Adam optimizer with a learning rate of 0.0001, five batch size and 500 training epochs. We evaluate the model using 10-fold cross validation.

\section{Experiment}

In this section, we review our dataset, the pre-processing steps, the hardware environment used for our experiments, and the key result of W-Net. In addition, we compare W-Net with ResNet~\cite{he2016deep}, a state-of-the-art approach, to demonstrate the effectiveness of our approach. We also highlight how one can use our model as a pre-trained model and fine-tune it using their own data. We show this in the context of public data use~\cite{rezatofighi2011automatic}. 
\vspace{-2mm}

\subsection{Dataset}

\begin{figure}[!t]
	\centering
	\includegraphics[width=0.47\textwidth]{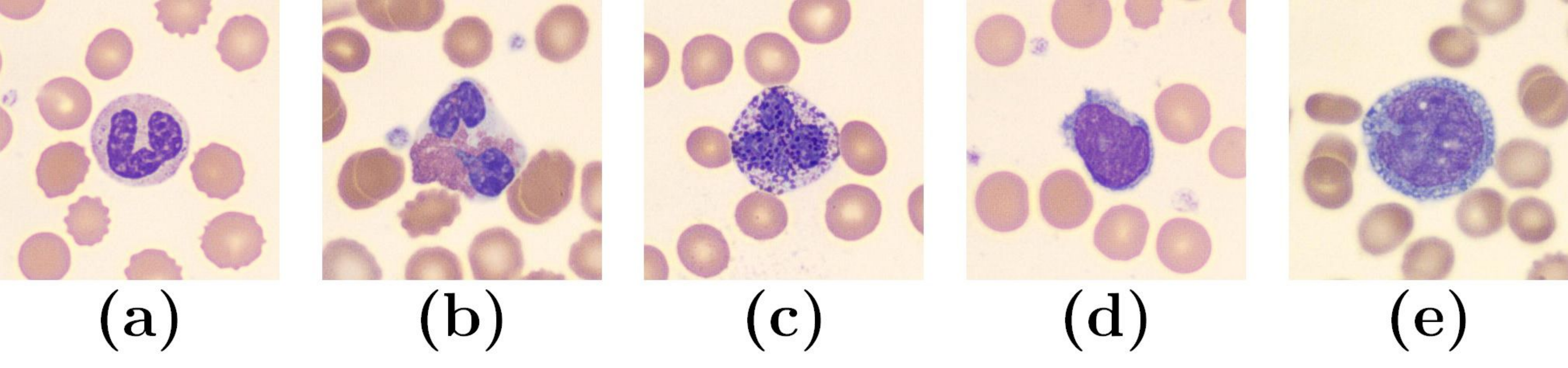}\vspace{-3mm}
	\caption{Examples from our dataset: (a) neutrophil, (b) eosinophil, (c) basophil, (d) lymphocyte, and (e) monocyte. Our dataset consists of 6,562 WBC images for five classes. Each image is 360$\times$361$\times$ 3 (i.e., 3 channels, RGB).
	}~\label{fig:wbc_original_imgs}\vspace{-7mm}
\end{figure}

\begin{table}[!t]
	\caption{The number of neutrophil, eosinophil, basophil, lymphocyte and monocyte samples in the dataset.}
	\vspace{2mm}
	\renewcommand{\arraystretch}{1.2}
	\label{table:the_number_of_images_of_each_class}
	\centering
	\begin{tabular}{cccccc}
		\toprule
		& NE & EO & BA & LY & MO \\
		\hline
		The \# of Imgs. & 2,006 & 1,310 & 377 & 1,676 & 1,193 \\
		Distribution & 30\% & 20\% & 6\% & 26\% & 18\% \\
		\bottomrule
	\end{tabular}\vspace{-5mm}
\end{table}

Our dataset has 6,562 real WBC images for the  five aforementioned classes (neutrophil, eosinophil, basophil, lymphocyte and monocyte), which were provided by The Catholic University of Korea (The CUK). The images were approved by the Institutional Review Board (IRB) of The CUK~\cite{CUKIRB}. The images were shot by Sysmex DI-60 machine~\cite{SysmexDI60}. Figure~\ref{fig:wbc_original_imgs} shows samples from the dataset used for this work. The size of each image is 360$\times$361$\times$3 (3 channels, RGB) pixels. Table~\ref{table:the_number_of_images_of_each_class} shows the number of images per class: 2,006 neutrophils images, 1,310 eosinophils images, 377 basophils images, 1,676 lymphocytes images and 1,193 monocytes images. Class distribution is 30\%, 20\%, 6\%, 26\% and 18\% in the dataset.

\vspace{-2mm}\subsection{Pre-processing}
Before training the model, images are pre-processed using 3 steps: 1) cropping, 2) re-sizing, and 3) normalizing. To extract features of WBC from the images, we cut the top, bottom, left and right sides of the image by 80, 81, 80 and 80 pixels, respectively. As a result, we obtained cropped images of size 200$\times$200$\times$3. We then re-sized the images to 128$\times$128$\times$3 for properly fitting them on a GPU memory and for efficient processing. To reduce the heterogeneity of the RGB distribution in the images and to prevent overflow/underflow, normalization was applied. Figure~\ref{fig:w-net_network_img} shows an overview of the pre-processing.
\vspace{-2mm}

\subsection{Hardware Environment}

W-Net was implemented using Tensorflow~\cite{abadi2016tensorflow}, trained on 2 Nvidia GTX 1080 with 8 GB of memory each, and hosted on a machine running Ubuntu 18.04.1 LTS operating system and using a 3.4 GHz Inter (R) Core (TM) i7-6700 CPU with 32 GB of main memory (RAM).
\vspace{-2mm}

\subsection{Result of W-Net}

\begin{table*}[!t]
	\renewcommand{\arraystretch}{1}
	\centering
	\caption{The result of 10-fold cross validation of W-Net for classification accuracy. The average accuracy for five classes is 97\%}
	\vspace{2mm}
	\label{table:result_cv_w-net}
	{\footnotesize
	\scalebox{1}{
	\centering
	\begin{tabular}{p{0.12\linewidth}cccccccccccp{0.10\linewidth}}
		\toprule
		& Fold-0 & Fold-1 & Fold-2 & Fold-3 & Fold-4 & Fold-5 & Fold-6 & Fold-7 & Fold-8 & Fold-9 & Aver. Accuracy \\
		\hline
		Neutrophil & 100\% & 98\% & 96\% & 97\% & 100\% & 100\% & 100\% & 95\% & 100\% & 98\% & 98\% \\
		Eosinophil & 95\% & 99\% & 93\% & 99\% & 100\% & 98\% & 98\% & 98\% & 93\% & 100\% & 97\% \\
		Basophil & 92\% & 94\% & 100\% & 100\% & 97\% & 94\% & 94\% & 94\% & 94\% & 91\% & 95\% \\
		Lymphocyte & 99\% & 100\% & 95\% & 95\% & 98\% & 97\% & 97\% & 98\% & 97\% & 95\% & 97\% \\
		Monocyte & 96\% & 100\% & 98\% & 96\% & 97\% & 98\% & 91\% & 96\% & 99\% & 97\% & 97\% \\
		Aver. Accuracy & 96\% & 98\% & 96\% & 97\% & 98\% & 97\% & 96\% & 96\% & 97\% & 96\% & 97\% \\
		\bottomrule
	\end{tabular}}}\vspace{-4mm}
\end{table*}

Table~\ref{table:result_cv_w-net} shows the results of the 10-fold cross validation of W-Net for classification accuracy. 
\cib{1} For neutrophil, 1,800 images were used for training and 206 images was used for testing per fold, and average accuracy was 98\%. 
\cib{2} For  eosinophil, 1,179 images were used for training  and 131 images were used for testing per fold, and the average accuracy was 97\%. 
\cib{3} For  basophil, 340 images were used for training  and 37 images were used for testing per fold, and the average accuracy was 95\%. 
\cib{4} For lymphocyte, 1,509 images were used for training and 167 images were used for testing per fold, and the average accuracy was 97\%. 
\cib{5} For the monocyte, 1,074 images were used for training and 119 images were used for testing per fold, and the average accuracy was 97\%. Overall, and by considering all of the above, the average accuracy for all five classes was 97\%.

Even though the basophil class has a relatively small distribution in the dataset, still we were able to achieve 95\% average accuracy. Also we achieved 97\% average accuracy for the five classes, notwithstanding the imbalance. As a result, we can claim that the results of our model are much more effective than previous studies.
\vspace{-2mm}

\subsection{ResNet}
\vspace{0mm}
\begin{table*}[!t]
	\renewcommand{\arraystretch}{1}
	\centering
	\caption{The results of ResNet for classification using 10-fold cross validation. The average accuracy for five classes is 51\%.}
	\vspace{2mm}
	\label{table:result_cv_resnet}
	{\footnotesize
	\scalebox{1}{
	\centering
	\begin{tabular}{p{0.12\linewidth}cccccccccccp{0.10\linewidth}}
		\toprule
		& Fold-0 & Fold-1 & Fold-2 & Fold-3 & Fold-4 & Fold-5 & Fold-6 & Fold-7 & Fold-8 & Fold-9 & Aver. Accuracy \\
		\hline
		Neutrophil & 100\% & 0\% & 100\% & 99\% & 0\% & 0\% & 0\% & 100\% & 100\% & 0\% & 50\% \\
		Eosinophil & 0\% & 16\% & 90\% & 95\% & 1\% & 23\% & 98\% & 1\% & 95\% & 87\% & 51\% \\
		Basophil & 0\% & 26\% & 94\% & 100\% & 78\% & 5\% & 86\% & 10\% & 100\% & 56\% & 56\% \\
		Lymphocyte & 49\% & 94\% & 5\% & 81\% & 67\% & 100\% & 0\% & 54\% & 33\% & 0\% & 48\% \\
		Monocyte & 1\% & 50\% & 100\% & 100\% & 1\% & 24\% & 100\% & 1\% & 23\% & 100\% & 50\% \\
		Aver. Accuracy & 30\% & 37\% & 78\% & 95\% & 29\% & 30\% & 57\% & 33\% & 70\% & 49\% & 51\% \\
		\bottomrule
	\end{tabular}}}\vspace{-3mm}
\end{table*}

In a deep learning network, if we use too many layers, we may get better results, but also the vanishing gradient problem may occur. ResNet (\underline{Res}idual neural \underline{Net}works)~\cite{he2016deep} can solve this problem by utilizing skip connections or short-cuts to jump over layers. Typical ResNet models are implemented with double or triple layer skips. ResNet is a CNN widely used in image processing.

In this section, we compared W-Net with ResNet to demonstrate the effectiveness of W-Net in WBC image classification. We trained a new model with ResNet50 network (50 Conv. layers) which has ReLU, softmax loss function and momentum optimizer. The best hyperparameters that are found were as follows: learning rate = 0.001, decay = 0.0001, momentum = 0.9, batch size = 32 and training epochs = 50. Also we used same dataset, pre-processing (except for the image size, we re-sized the images to 224 $\times$ 224 $\times$ 3 for ResNet), hardware environment and 10-fold cross validation splits were the same with W-Net experiment. The network has a total of 23,544,837 trainable parameters. Table~\ref{table:result_cv_resnet} shows the result of 10-fold cross validation of ResNet for classification accuracy. For the neutrophil 50\%, eosinophil 51\%, basophil 56\%, lymphocyte 48\% and monocyte 50\% 10-fold cross validation accuracies are obtained. The average accuracy for five classes is 51\%.

In the 10-fold cross validation evaluation of W-Net, minimum average accuracy is 91\% (basophil, Fold-9) and maximum average accuracy is 100\%. However, in the case of ResNet the variance between the folds is from 0\% to 100\% resulting 51\% 10-fold average accuracy. This means that deep networks may be not efficient for WBC image classification. As a result, we can claim that W-Net which has five layers (three Conv. and two FC.) is more effective than ResNet in WBC image classification area.
\vspace{-2mm}

\subsection{Further Training with Public Data}

\begin{figure}[!t]
	\centering
	\includegraphics[scale=0.37]{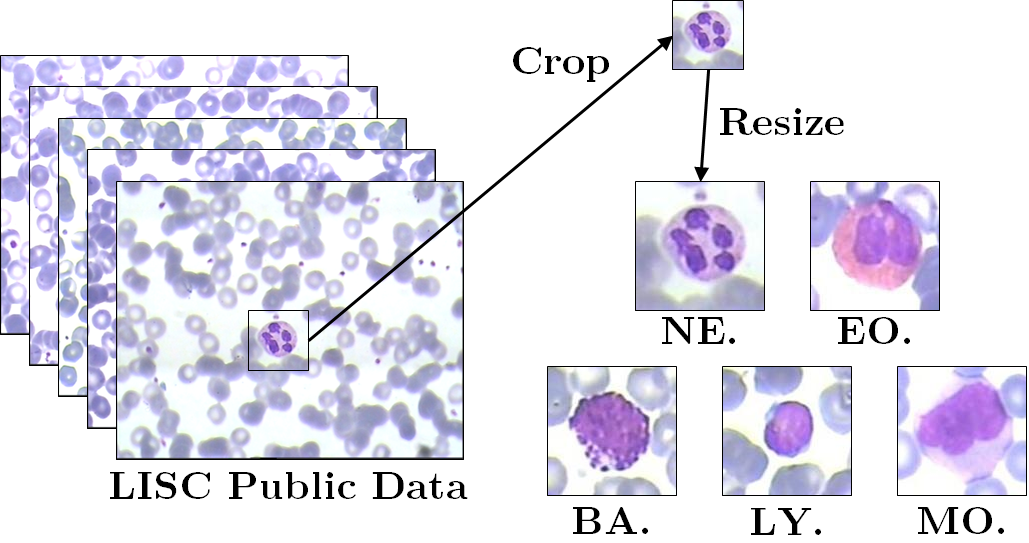}
	\caption{
		LISC public data.
	}~\label{fig:lisc_public_data}\vspace{-8mm}
\end{figure}

\begin{table*}[!t]
	\renewcommand{\arraystretch}{1}
	\centering
	\caption{The result of the first model trained using LISC public data from scratch. The average accuracy for five classes is 91\%.}
	\vspace{2mm}
	\label{table:result_public_data_from_scratch}
	{\footnotesize
	\scalebox{1}{
	\centering
	\begin{tabular}{p{0.12\linewidth}cccccccccccp{0.10\linewidth}}
		\toprule
		& Fold-0 & Fold-1 & Fold-2 & Fold-3 & Fold-4 & Fold-5 & Fold-6 & Fold-7 & Fold-8 & Fold-9 & Aver. Accuracy \\
		\hline
		Neutrophil & 33\% & 83\% & 100\% & 100\% & 83\% & 83\% & 100\% & 80\% & 80\% & 100\% & 84\% \\
		Eosinophil & 100\% & 100\% & 100\% & 100\% & 100\% & 100\% & 100\% & 100\% & 100\% & 100\% & 100\% \\
		Basophil & 50\% & 83\% & 100\% & 100\% & 100\% & 100\% & 100\% & 100\% & 80\% & 80\% & 89\% \\
		Lymphocyte & 100\% & 100\% & 100\% & 100\% & 83\% & 100\% & 100\% & 80\% & 100\% & 100\% & 96\% \\
		Monocyte & 60\% & 100\% & 20\% & 80\% & 100\% & 100\% & 100\% & 100\% & 100\% & 100\% & 86\% \\
		Aver. Accuracy & 69\% & 93\% & 84\% & 96\% & 93\% & 97\% & 100\% & 92\% & 92\% & 96\% & 91\% \\
		\bottomrule
	\end{tabular}}}\vspace{-3mm}
\end{table*}

\begin{table*}[!t]
	\renewcommand{\arraystretch}{1}
	\centering
	\caption{The result of the second model was initially trained using our dataset which is our W-Net model and then further trained using LISC public data. The average accuracy for five classes is 96\%.}
	\vspace{2mm}
	\label{table:result_public_data_more_trained}
	{\footnotesize
	\scalebox{1}{
	\centering
	\begin{tabular}{p{0.12\linewidth}cccccccccccp{0.10\linewidth}}
		\toprule
		& Fold-0 & Fold-1 & Fold-2 & Fold-3 & Fold-4 & Fold-5 & Fold-6 & Fold-7 & Fold-8 & Fold-9 & Aver. Accuracy \\
		\hline
		Neutrophil & 100\% & 100\% & 100\% & 100\% & 100\% & 100\% & 100\% & 100\% & 100\% & 100\% & 100\% \\
		Eosinophil & 100\% & 100\% & 100\% & 100\% & 100\% & 100\% & 100\% & 75\% & 100\% & 100\% & 98\% \\
		Basophil & 100\% & 100\% & 100\% & 100\% & 100\% & 100\% & 100\% & 100\% & 80\% & 80\% & 96\% \\
		Lymphocyte & 100\% & 100\% & 100\% & 100\% & 100\% & 83\% & 100\% & 100\% & 100\% & 100\% & 98\% \\
		Monocyte & 80\% & 100\% & 20\% & 100\% & 80\% & 100\% & 100\% & 100\% & 100\% & 100\% & 88\% \\
		Aver. Accuracy & 96\% & 100\% & 84\% & 100\% & 96\% & 97\% & 100\% & 95\% & 96\% & 96\% & 96\% \\
		\bottomrule
	\end{tabular}}}\vspace{-4mm}
\end{table*}

The LISC public data~\cite{rezatofighi2011automatic} provides WBC image samples that were taken from peripheral blood of 8 normal subjects. The images contain 720$\times$576 pixels. The images are classified by a hematologist into normal leukocytes: neutrophils, eosinophils, basophils, lymphocytes and monocytes. For pre-processing the public data, we cropped the WBC images (nucleus and cytoplasm regions) in the original images, and then re-sized the images to 128$\times$128 $\times$3 for training as shown in Figure~\ref{fig:lisc_public_data}. We used a total of 254 WBC images as our dataset: 56 neutrophil images, 39 eosinophil images, 55 basophil images, 56 lymphocyte images and 48 monocyte images.

We conducted an experiment using the LISC public data to show how other researchers can benefit from our trained W-Net model. Specifically, one can further tune/train our W-Net model for better performance. As such, we implemented two models of the same W-Net architecture: \cib{1} a model  trained using only LISC public data from scratch, and \cib{2} a model initially trained using our dataset using W-Net model and then further trained using LISC public data. 

Except the training epochs, the hyperparameters of both models were identical. The first model was trained 4,000 epochs (254$\times$4,000/5 iterations) on the public data. The second model was trained for 500 epochs (6,562$\times$500/5 iterations) on our dataset and then trained for 4,000 epochs (254$\times$4,000/5 iterations) on the public data. We used 128$\times$128$\times$3 public images, the same normalization, hardware environment, and 10-fold cross validation as in W-Net experiment. The network has a total of 16,806,949 trainable parameters. Table~\ref{table:result_public_data_from_scratch} shows the result of the first model using LISC public data. The average accuracy for five classes was 91\%. Table~\ref{table:result_public_data_more_trained} shows the result of the second model, where the average accuracy for five classes was 96\%.

In comparing both, the second model shows better results. As such, we can claim that our model can help other researchers in the field of WBC image classification, and we released our model on our github~\cite{WNetModel}.

\section{Conclusion}

The analysis of WBC count and types is essential for diagnosing diseases. Even though there are several methods for detecting and counting WBCs from microscopic images of a blood smear, the recognition of the five types of WBCs (namely, neutrophils, eosinophils, basophils, lymphocytes and monocytes) is still a challenge in real-life applications, which we addressed in this work. The rapid advancements in the field of computer vision and machine learning have provided feasible solutions to accurate classification tasks in many domains. This work proposes W-Net, a CNN-based architecture, to enable accurate classification of the five WBC types. We evaluated the proposed architecture on a real-life dataset and addressed several challenges such as the transfer learning property and the class imbalance. W-Net achieved an average classification accuracy of 97\%. Moreover, we compared the results of W-Net and ResNet architectures to show the superiority of W-Net over other architecture. 
\vspace{-2mm}

\subsection{Acknowledgement} This research was supported by the Global Research Laboratory (GRL) Program of the National Research Foundation (NRF) funded by Ministry of Science, ICT (Information and Communication Technologies) and Future Planning (NRF-2016K1A1A2912757), and by the Technology Innovation Program (No: 10049771, Development of Highly-Specialized Platform for IVD Medical Devices, and No: 10059106, Development of a smart white blood cell image analyzer with 60t/h throughput and sub-$\mu$m imaging device, based on Manual Review Center with less than 1\% analysis error) funded by the Ministry of Trade, Industry \& Energy, Republic of Korea. DaeHun Nyang is the corresponding author.
\vspace{-2mm}


\balance

\bibliographystyle{aaai}
\bibliography{bibilo.bib}

\end{document}